\newcommand{\upcite}[1]{\textsuperscript{\textsuperscript{\cite{#1}}}}
\documentclass[twocolumn,showpacs,preprintnumbers,amsmath,amssymb]{revtex4}

\usepackage{amsmath}
\usepackage{amssymb}
\usepackage{graphicx}

\begin{document}

\title{Proof-of-principle experimental demonstration of quantum secure imaging based on quantum key distribution}
\author{Yi-Bo Zhao$^{1,2}$, \ Wan-Li Zhang$^{1,2}$, \ Dong Wang$^{1,2}$, \ Xiao-Tian Song$^{1,2}$,\ Liang-Jiang Zhou$^{1,2,3,}$\footnote{Corresponding author: E-mail: ljzhou@mail.ie.ac.cn}, and \ Chi-Biao Ding$^{1,2,3}$}
\address{$^{1}${\textsl{National Key Laboratory of Microwave Imaging, Beijing} 100190, \textsl{China}}\\  
$^{2}${\textsl{Aerospace Information Research Institute, Chinese Academy of Sciences, Beijing} 100190, \textsl{China}}\\ 
$^{3}${\textsl{University of Chinese Academy of Sciences, Beijing} 100049, \textsl{China}}}

\date{\today}

\begin{abstract}
We present a quantum secure imaging (QSI) scheme based on the phase encoding and weak$+$vacuum decoy-state BB84 protocol of quantum key distribution (QKD). It allows us to implement a computational ghost imaging (CGI) system with more simplified equipment and reconstructed algorithm by using a digital micro-mirror device (DMD) to preset the specific spatial distribution of the light intensity. What's more, the quantum bit error rate (QBER) and the secure key rate analytical functions of QKD are used to see through the intercept-resend jamming attacks and ensure the authenticity of the imaging information. In the experiment, we obtained the image of the object quickly and efficiently by measuring the signal photon counts with single-photon detector (SPD), and achieved a secure key rate of 571.0 bps and a secure QBER of 3.99$\%$, which is well below the lower bound of QBER of 14.51$\%$. Besides, Our imaging system uses a laser with invisible wavelength of 1550 nm, whose intensity is low as single-photon, that can realize weak-light imaging and is immune to the stray light or air turbulence, thus it will become a better choice for quantum security radar against intercept-resend jamming attacks.
\end{abstract}

\maketitle

\section{Introduction}
How to enhance the anti-jamming ability of radar is an urgent problem to be solved in the traditional radar countermeasures technology.\upcite{a,b} In the last two decades, with the hot research and development of the techniques of quantum communication,\upcite{e,f,g} quantum imaging,\upcite{h,i,j,k,l,ab,ac,ad,ae,af,ag,ah,m,n,o,p,q} and quantum radar,\upcite{r,s} a series of results have been achieved, which provides a solution for improving the anti-jamming ability of radar.

In 1995, quantum imaging was first observed,\upcite{h} also known as ghost imaging (GI). Entangled photon pairs produced by spontaneous parametric down-conversion were used, while one of the photons passing through the object was collected by a photon counter or a bucket detector with no spatial resolution (called the signal beam or object beam), its twin photon was detected by a multipixel detector [e.g., a charge-coupled device (CCD) camera] without ever passing through the object (called the reference beam). Nevertheless, by correlating the intensities measured by the bucket detector with the intensities of each pixel in the multipixel detector, an image of the object is reconstructed.\upcite{h} In a long time, entanglement swapping is considered as the necessary condition for GI.\upcite{t} However, from 2002 GI were implemented by classical light source,\upcite{k,l,m,ab,ac,ad,ae,af,ag,ah} especially the groups of Wang Kaige\upcite{ab}, Wu Ling-an\upcite{ac} and Han shensheng\upcite{ad,ae} accomplished a large numbers of GI experiments based on pseudo-thermal or real thermal light source, which proved that entanglement is not the one and only way to achieve GI. After then, computational ghost imaging (CGI) was proposed\upcite{n} and verified by experiments,\upcite{o,p} which could be deployed with only one optical beam and one detector with no spatial resolution by replacing the reference beam with a controllable and pre-modulated light source.\upcite{q} The greatly simplified imaging system reduces the number of preset patterns and the imaging computational complexity, and then raises the imaging efficiency.\upcite{q} Based on the above, CGI and quantum secure imaging (QSI) could be applied to radar imaging faultlessly. Ref. \cite{j} introduced an anti-jamming quantum radar imaging (QRI) that utilized the BB84 protocol quantum key distribution (QKD) based on polarization encoding, which could verify the authenticity of imaging information by detecting the consistency of photon polarization states between the sender and the receiver. Up to now, the BB84 protocol is the most mature and widely used quantum key distribution protocol and its theoretical security has been fully proved,\upcite{u,v} Therefore, intercept-resend attacks is exposed immediately and the authenticity of the imaging information can be guaranteed through combining the CGI and QKD on radar technology. In addition, the quantum imaging technology has many advantages of higher resolution, higher contrast, higher signal-to-noise ratio (SNR), and less influence of atmospheric turbulence\upcite{i} compared with the traditional imaging. QSI will has great application value in the fields of aerial detection, military reconnaissance and battlefield imaging in the future.

Inspired by ref. \cite{j}, we present a QSI scheme combining the phase-encoding BB84 QKD and CGI technology. Unlike the original scheme, our scheme adopt phase-encoding QKD, in which the relative phase is rather stable against disturbance. Moreover, owing to the fact that perfect single-photon source is still unavailable, a pulsed laser source with Poisson distribution is used in the QKD system. By combing the decoy-state method, we can estimate the contribution of multi-photon part to obtain the lower bound of QBER under the intercept-resend jamming attack, which is tighter than the origin one. In our experiment, A DMD with quick response speed is used to preset the specific spatial distribution of the light intensity, and single-photon detector (SPD) with quick response speed is used to measure the total light intensity, eventually the image of the object is retrieved through calculating the intensity distribution and the total intensity of light. Comparing with the traditional ghost imaging (TGI), QSI has been greatly improved in the data sampling rate and reduced in the imaging algorithm complexity and shortened the imaging time. More importantly, the quantum bit error rate (QBER) and the secure key rate of QKD are obtained in the experiment to see through the intercept-resend jamming attacks and ensure the authenticity of the imaging information.

\section{Theory and method}
\begin{figure}
  \centering
  \includegraphics[width=1\columnwidth]{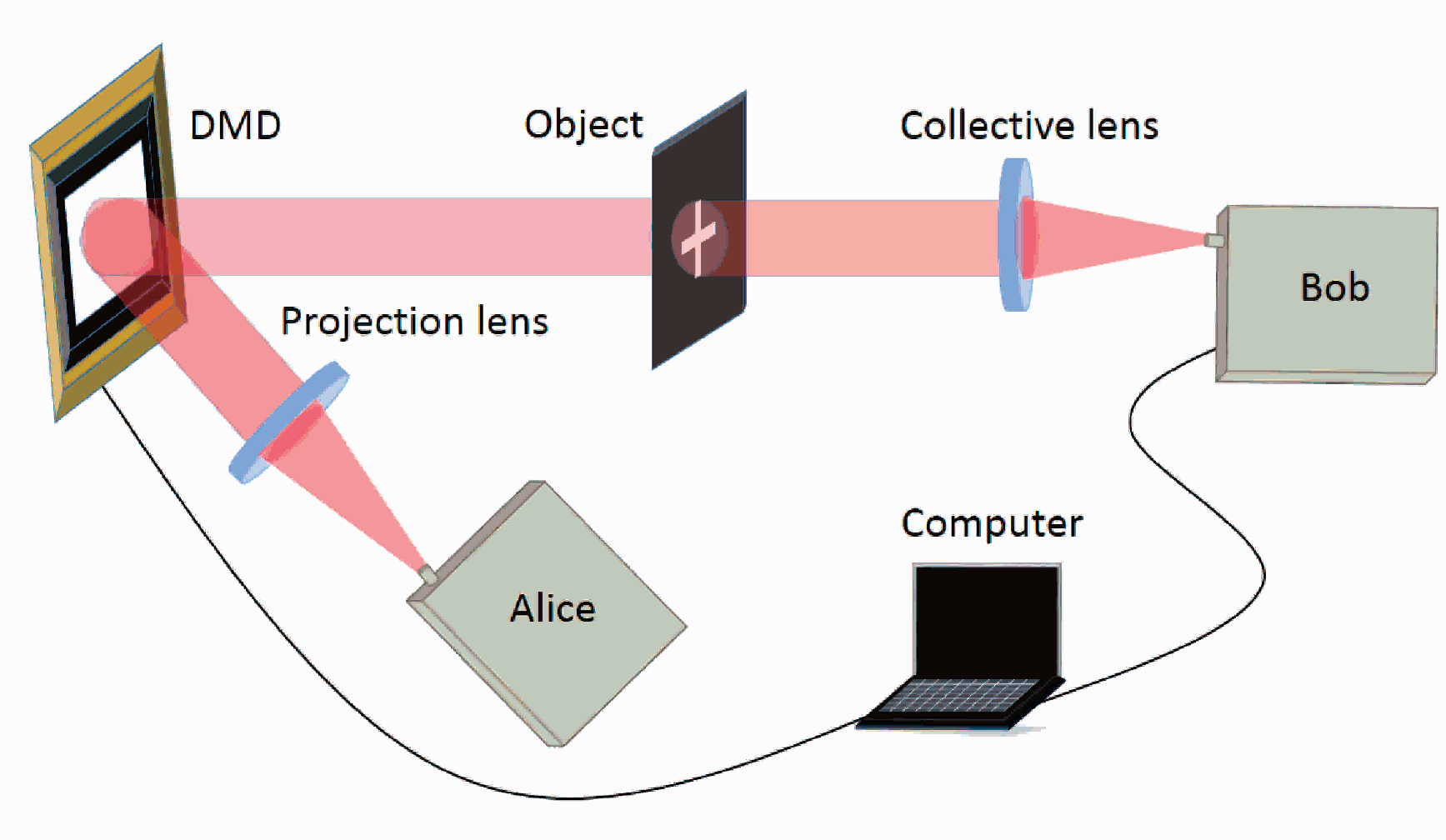}\\
  \caption{(color online) Schematic of the quantum secure imaging system. Polarized single-photon from Alice, passes through the beam expanding lens, is modulated into specific spatial intensity distribution by the DMD, then transmits through the object, finally is coupled into Bob and detected by SPDs. The image of the object is reconstructed and the QBER is monitored to estimate whether the imaging process is safe and real.The angle of reflection from the DMD is exaggerated in the figure for clarity but is about 24° in reality.}
  \label{QSI}
\end{figure}

The QSI system contains the imaging system and the QKD system, as shown in Fig. \ref{QSI}, which mainly includes QKD sender (Alice), QKD receiver (Bob), and the spatial light path of CGI in more details. In our scheme, Alice sends out strongly attenuated coherent states with key bits information, which are encoded into the relative phase of a pair of coherent states shifted in time that generated by an asymmetric interferometer. These photons passes through the beam expanding lens, the obtained speckle is then modulated into a specific spatial distribution of intensity with a DMD. The pulse with certain intensity distribution illuminates the object, and is coupled into the receiver, perform the decoding and detecting procedure. Finally we can reconstruct the image of object with the detected counts. It should be stressed that, in QKD procedure the key bits information is encoded into the relative phase between the two time-bin mode of signal pulses, which is rather stable during the display time of one frame of the DMD, since the modulation of DMD just changes the global phase of the whole two-time-bin pulse rather than the relative phase between them. Therefore, the pulses passing through the imaging optical path and arrive at Bob without changing its coding information, and can be further correctly decoded by Bob. Compared with the conventional CGI system, the major difference is that the light source and detector are replaced by a QKD sender and receiver in our QSI scheme, with which the security of imaging can be promoted. As it is necessary to measure the photon counts and QBER during the display time of every pattern loaded in the DMD, the trigger signals of the laser, the DMD and the SPDs should be synchronized. The image of the object is reconstructed through calculating the pre-modulated intensity distribution of the DMD and the measured total intensity of SPDs, while the QBER is monitored to estimate whether the imaging process is safe and real.

\subsection{Quantum key distribution system}
\begin{figure}
  \centering
  \includegraphics[width=1\columnwidth]{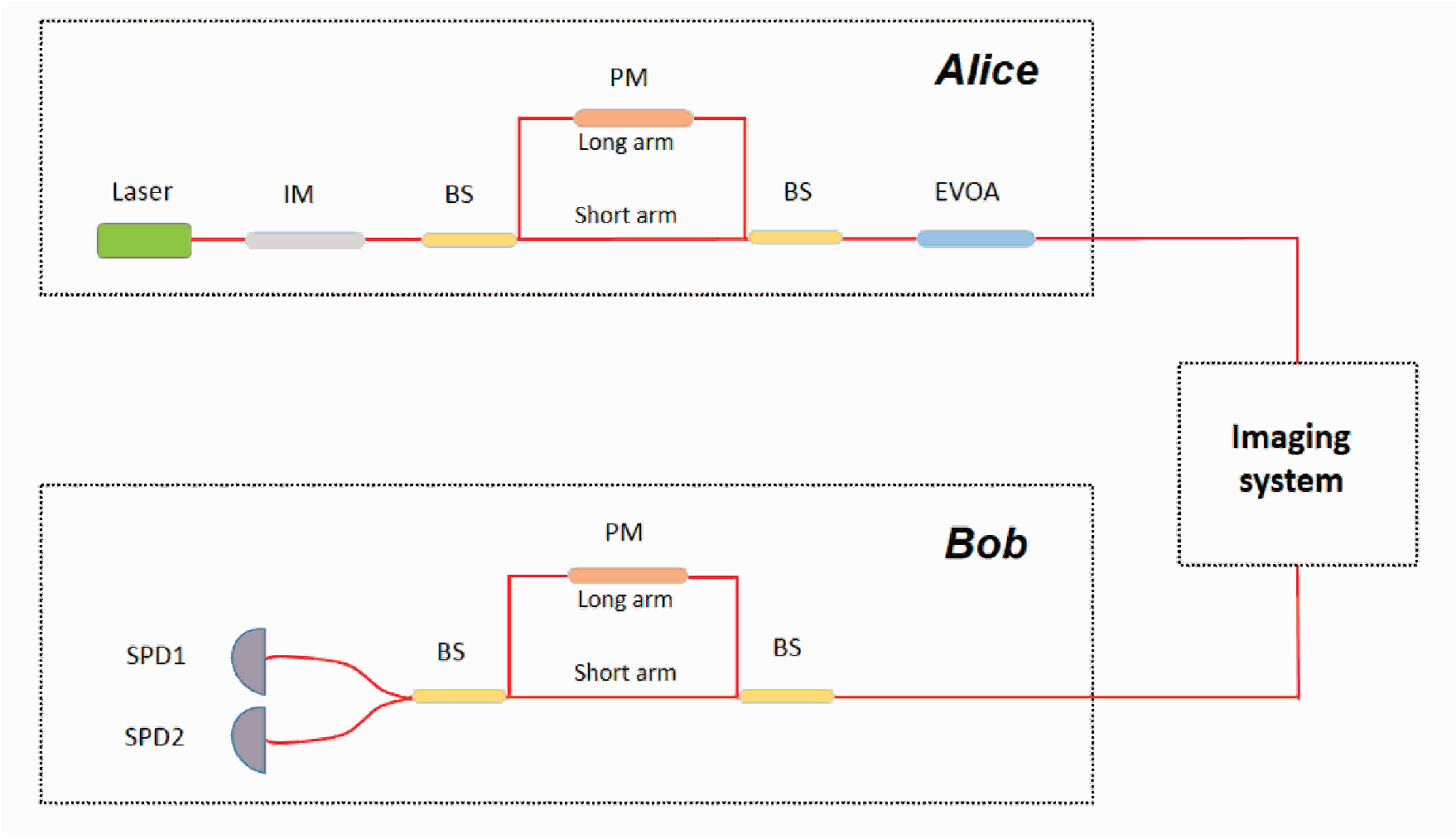}\\
  \caption{(color online) Schematic of the phase encoding and weak$+$vacuum decoy-state BB84 protocol QKD. IM: Intensity modulator; BS: Beam splitter; PM: Phase modulator; EVOA: Electronic variable optical attenuator; SPD: Single-photon detector. The signal and decoy pulses are produced by IM. Two identical UMZIs in Alice and Bob play roles of encoder. An EVOA is used to attenuate the intensity of the laser pulses to the single-photon level.} \label{QKD}
\end{figure}

Quantum key distribution technology is one of the most mature and commonly used communication methods in quantum communication, as well as the BB84 protocol in QKD protocol, whose theoretical security have been fully proved.\upcite{u,v} So, we adopt the scheme based on the phase encoding BB84 protocol of QKD.\upcite{e,f} Our QKD system is shown in Fig. \ref{QKD}. It is composed of two parts of Alice and Bob. Alice uses a pulsed laser to produce coherent pulses. An intensity modulator is used to produce signal and decoy pulses of differing intensities. The vacuum pulses is produced by omitting trigger signal to the laser. An unbalanced Mach-Zehnder interferometer (UMZI) with a phase modulator (PM) in the long arm plays a role of encoder, while there is an identical UMZI in Bob.\upcite{e} SPDs are deployed by Bob to detect the interferometric outcomes from the interferometer. Moreover, an electronic variable optical attenuator (EVOA) is used to attenuate the intensity of the laser pulses to the single-photon level before sending out from the sender.\upcite{u} In the experiment, the intensities of signal and decoy pulses are set as $\mu$ and $\nu$ to perform the weak$+$vacuum decoy-state BB84 QKD protocol.\upcite{aa}

According to the phase encoding BB84 protocol, the relative phase on Alice is selected randomly from four values $0, \pi/2, \pi$ and $3\pi/2$, where $0,\pi$ are in the Z basis, and $\pi/2,3\pi/2$ are in the X basis. Bob measures the incoming quantum states from Alice also in Z or X basis randomly. Whenever Alice and Bob use the same basis, the two SPDs will get correct results with a small QBER. However, when they use different bases, the two SPDs will get an uncertain result together. Alice then reveals only whether or not the state in which she encoded that qubit is compatible with the basis announced by Bob. If the state is compatible, they keep the bit; if not, they disregard it. In this way, fifty percent of the bit string is discarded. This shorter key obtained after basis reconciliation is called the sifted key.

\subsection{Intercept-resend eavesdropping strategy}
In the practical QKD system, Alice uses the weak coherent source (WCS) together with the decoy-state method to substitute the unavailable single-photon source. The photon number of each pulse follows Poisson distribution. The probability of finding an \emph{n}-photon state in a pulse is given by ${P_n}\left( \mu  \right) = \frac{{{e^{ - \mu }}{\mu ^n}}}{{n!}}$, where $\mu$ is the mean photon number. With the phases randomization, the states sent by Alice are described as
\begin{equation}
{\rho _k} = \sum\limits_{n = 0}^\infty  {\frac{{{e^{ - \mu }}{\mu ^n}}}{{n!}}\left| {{n_k}} \right\rangle \left\langle {{n_k}} \right|} ,
\end{equation}
where the states $\left| {{n_k}} \right\rangle$ denote Fock states with $n$ photons in one of the four BB84 polarization states, which are labeled with the index $k$, with $k=0,\cdots,3$ (correspond to $0, \pi/2, \pi$ and $3\pi/2$, respectively).

In the case of quantum secure imaging or quantum radar, the potential attacker usually adopt the intercept-resend strategy to forge the target. Since the QKD system is applied to enhance its security, here we just consider the intercept-resend quantum eavesdropping strategy that proposed by M. Curty et al.\upcite{bb} Eve's attack strategy can be decomposed into three steps. In the first step, Eve obtains the total photon number $n$ of each pulse sent by Alice via a quantum non-demolition (QND) measurement without introducing any errors. Then she performs a filter operation on the state $\left| {{n_k}} \right\rangle$ with the intention to make them distinguishable with some finite probability. Finally, in the third step, Eve measures out each filtered state with the so-called square-root measurement (SRM). This measurement gives her the minimum value of the error probability when distinguishing symmetric states. After deciding which state was used by Alice, Eve needs to prepare a new signal in the state identified and give it to Bob. The error rate introduced by Eve with this eavesdropping strategy for a signal state containing $n$ photon is given by

\begin{equation}\label{en}
{e_n} = \frac{1}{2} - \frac{{\sum\limits_{l,m = 0}^1 {\left| {{\gamma _{2l}}\left( n \right){\gamma _{2m + 1}}\left( n \right)} \right|} }}{{2\sum\limits_{j = 0}^3 {{{\left| {{\gamma _j}\left( n \right)} \right|}^2}} }},
\end{equation}
where $\gamma_j(n)=\alpha_j(n)c_j(n)$, with coefficients $\alpha_j(n)$ satisfying ${\left| {{\alpha _j}\left( n \right)} \right|^2} \leqslant 1$ for all $j = 0,\cdots,3$ and for all $n\ge0$. And the coefficients $c_j(n)$ satisfy $\sum\nolimits_j {{{\left| {{c_j}\left( n \right)} \right|}^2}}  = 1$. The exact values of the coefficients ${\left| {{c_j}\left( n \right)} \right|}$ can be obtained explicitly as follows by using the overlaps of the four states $\left| {{n_k}} \right\rangle$, with $k=0,\cdots,3.$
\begin{equation}\label{cj}
\begin{array}{l}
\left| {{c_0}\left( n \right)} \right| = \sqrt {\frac{1}{4} + {2^{ - \left( {1 + {n \mathord{\left/
 {\vphantom {n 2}} \right.
 \kern-\nulldelimiterspace} 2}} \right)}}\cos \left( {\frac{\pi }{4}n} \right)} ,\\
\left| {{c_1}\left( n \right)} \right| = \sqrt {\frac{1}{4} + {2^{ - \left( {1 + {n \mathord{\left/
 {\vphantom {n 2}} \right.
 \kern-\nulldelimiterspace} 2}} \right)}}\sin \left( {\frac{\pi }{4}n} \right)} ,\\
\left| {{c_2}\left( n \right)} \right| = \sqrt {\frac{1}{4} - {2^{ - \left( {1 + {n \mathord{\left/
 {\vphantom {n 2}} \right.
 \kern-\nulldelimiterspace} 2}} \right)}}\cos \left( {\frac{\pi }{4}n} \right)} ,\\
\left| {{c_3}\left( n \right)} \right| = \sqrt {\frac{1}{4} - {2^{ - \left( {1 + {n \mathord{\left/
 {\vphantom {n 2}} \right.
 \kern-\nulldelimiterspace} 2}} \right)}}\sin \left( {\frac{\pi }{4}n} \right)} ,
\end{array}
\end{equation}

For different states $\left| {{n_k}}\right\rangle$ we can obtain the minimum error rate with Eq. (\ref{en}) and Eq. (\ref{cj}). In the following we calculate the case that the signal sent by Alice contains only $0< n \leqslant 2$ photons. In the case of single photon state $\left| {{1_k}}\right\rangle$, that is $n=1$, from Eq. (\ref{cj}) we have $\left| {{c_0}\left( 1 \right)} \right|= \left| {{c_1}\left( 1 \right)} \right| ={1 \mathord{\left/ {\vphantom {1 {\sqrt 2 }}} \right. \kern-\nulldelimiterspace} {\sqrt 2 }}$ and $\left| {{c_2}\left( 1 \right)} \right|= \left| {{c_3}\left( 1 \right)} \right| =0$. From Eq. (\ref{en}) we obtain that, in this case, the partial error rate $e_1$ is given by
\begin{equation}\label{e1}
{e_1} = \frac{1}{2} - \frac{1}{2}\frac{{\left| {{\alpha _0}\left( 1 \right){\alpha _1}\left( 1 \right)} \right|}}{{{{\left| {{\alpha _0}\left( 1 \right)} \right|}^2} + {{\left| {{\alpha _1}\left( 1 \right)} \right|}^2}}}
\end{equation}
It is easy to get that the minimum value of $e_1 =0.25 $ for all ${\left| {{\alpha _j}\left( 1 \right)} \right|^2} \leqslant 1$.

The analysis for the case $n=2$ is similar. The coefficients $c_j(2)$ of the signal states $\left| {{2_k}}\right\rangle$ are of the form $\left| {{c_0}\left( 2 \right)} \right|= \left| {{c_2}\left( 2 \right)} \right| = {1 \mathord{\left/  {\vphantom {1 2}} \right.  \kern-\nulldelimiterspace} 2}$, $\left| {{c_1}\left( 2 \right)} \right| ={1 \mathord{\left/ {\vphantom {1 {\sqrt 2 }}} \right. \kern-\nulldelimiterspace} {\sqrt 2 }}$ and $\left| {{c_3}\left( 2 \right)} \right|=0$, respectively. The partial error rate $e_2$ is given by
\begin{equation}\label{e2}
{e_2} = \frac{1}{2} - \frac{1}{{\sqrt 2 }}\frac{{\left| {{\alpha _0}\left( 2 \right){\alpha _1}\left( 2 \right)} \right| + \left| {{\alpha _1}\left( 2 \right){\alpha _2}\left( 2 \right)} \right|}}{{{{\left| {{\alpha _0}\left( 2 \right)} \right|}^2} + 2{{\left| {{\alpha _1}\left( 2 \right)} \right|}^2} + {{\left| {{\alpha _2}\left( 2 \right)} \right|}^2}}}.
\end{equation}
Then we can obtain the minimum value of $e_2={{\left( {2 - \sqrt 2 } \right)} \mathord{\left/ {\vphantom {{\left( {2 - \sqrt 2 } \right)} 4}} \right. \kern-\nulldelimiterspace} 4}\approx 0.146$.

In order to give the lower bound of total error rate of signal state at the present of intercept-resend attack, we adopt model of decoy-state method to estimate the yield of photon pulses.\upcite{aa} In the weak$+$vacuum decoy-state method, Alice randomly prepare the signal, decoy and vacuum state with intensity of $\mu, \nu$ and 0, respectively. For $0 < \nu < \mu \leqslant 1$, the following inequalities hold for all $n \geqslant 3$,
\begin{align}\label{condition}
\frac{{{P_n}\left( \mu  \right)}}{{{P_n}\left( \nu  \right)}}  \ge \frac{{{P_3}\left( \mu  \right)}}{{{P_3}\left( \nu  \right)}} ,
{P_1}\left( \mu  \right)  > {P_1}\left( \nu  \right)   ,
{P_2}\left( \mu  \right)  > {P_2}\left( \nu  \right) .
\end{align}

The overall gains of signal and decoy state are given by
\begin{align}\label{gain}
{Q_\mu } = \sum\limits_{n = 0}^\infty  {{P_n}\left( \mu  \right){Y_n}} \\ \nonumber
{Q_\nu } = \sum\limits_{n = 0}^\infty  {{P_n}\left( \nu  \right){Y_n}}
\end{align}
where $Y_n$ is the yield of an $n$-photon state, i.e., the conditional probability of a detection event at Bob's side given that Alice sends out an $i$-photon state.

By using the inequalities (\ref{condition}) and Eq. (\ref{gain}), the joint contribution of one-photon and two-photon component can be deduced as follows
\begin{equation}
\begin{aligned}
&{P_1}\left( \nu  \right){Y_1} + {P_2}\left( \nu  \right){Y_2}  \\ 
\geqslant & \frac{{{P_3}\left( \mu  \right){Q_\nu } - {P_3}\left( \nu  \right){Q_\mu } - \left[ {{P_0}\left( \nu  \right){P_3}\left( \mu  \right) - {P_0}\left( \mu  \right){P_3}\left( \nu  \right)} \right]{Y_0}}}{{{P_3}\left( \mu  \right) - {P_3}\left( \nu  \right)}},
\end{aligned}
\end{equation}

where $Y_0$ is the background rate, which includes the detector dark count and other background contributions such as the stray light from timing pulses.

The total error rate of decoy state when Eve performs the intercept-resend attack is given by
\begin{equation}
\begin{aligned}\label{et}
{E_{\nu}Q_{\nu}} &= \mathop \sum \limits_{i = 0}^\infty  {e_i}{P_i}\left( \nu  \right){Y_i} = {e_0}{P_0}\left( \nu  \right){Y_0}\\
&  +{e_1}{P_1}\left( \nu  \right){Y_1} + {e_2}{P_2}\left( \nu  \right){Y_2} +  \mathop \sum \limits_{i = 3}^\infty  {e_i}{P_i}\left( \nu  \right){Y_i}  \\ \nonumber
& \geqslant {P_0}\left( \nu  \right){Y_0}/2 + {e_2} \left[ {P_1}\left( \nu  \right){Y_1} + {P_2}\left( \nu  \right){Y_2} \right],
\end{aligned}
\end{equation}
where $e_i$ is the error rate of the $i$-photon state, and the error rate of the background $e_0 = \frac{1}{2}$.
Thus we can obtain the lower bound of decoy state total error rate
\begin{equation}
\begin{aligned}\label{etL}
E_{\nu}^L &= \frac{1}{Q_{\nu}} \{ \frac{1}{2}{P_0}\left( \nu  \right){Y_0} + {e_2}\frac{{{P_3}\left( \mu  \right){Q_\nu } - {P_3}\left( \nu  \right){Q_\mu } }}{{{P_3}\left( \mu  \right) - {P_3}\left( \nu  \right)}}\\
& -{e_2Y_0} \frac{\left[ {{P_0}\left( \nu  \right){P_3}\left( \mu  \right) - {P_0}\left( \mu  \right){P_3}\left( \nu  \right)} \right] }{{P_3}\left( \mu  \right) - {P_3}\left( \nu  \right)}  \}.
\end{aligned}
\end{equation}

Therefore, one can monitor the error rate of decoy state, when the images obtained from a signal with an error rate of our QSI system greater than $E_{\nu}^L$ cannot be considered secure and imply the existence of Eve. In addition, whether our QKD system obtain secure keys can also be a criteria of secure imaging.

\subsection{Computational quantum imaging system}
CGI is an improvement from TGI. A spatial light modulator (SLM) or a CCD is used to preset the specific spatial distribution of the light intensity. Disregarding the reference beam, so just a single-pixel detector is needed.\upcite{o} In this imaging scheme, taking the DMD to be a part of Alice. According to the security requirement of the decoy state BB84 protocol, the intensity of the laser pulses has to be lowered as single-photon level at the exit of the DMD by an EVOA.

As in our scheme, a DMD is used to preset  intensity patterns of speckles, and single-photon detectors are employed to measure the counts corresponding to each pattern. Suppose $N$ is the number of intensity patterns modulated by DMD, one can obtain the total count $B_i(1 \leqslant i \leqslant N)$ during the display time of the $i$th pattern, whose intensity distribution is denoted by $I_i(x, y)(1 \leqslant i \leqslant N)$. Then the image of the object can be obtained by summing the calculated intensities $I_i$ with the appropriate weights $B_i$, \upcite{o}
\begin{equation}\label{Oxy}
O\left( {x,y} \right) = \frac{1}{N}\mathop \sum \limits_{i = 1}^N \left( {{B_i} - \langle{B_i}\rangle} \right)\left[ {{I_i}\left( {x,y} \right) - \langle{ {I_i}\left( {x,y} \right)}\rangle} \right]\;
\end{equation}
where
\begin{equation}\label{Bi}
\langle{B_i}\rangle = \frac{1}{N}\mathop \sum \limits_{i = 1}^N {B_i}
\end{equation}
is the mean value of $B_i$, and $B_i$ meets the following formula,
\begin{equation}\label{Bi1}
{B_i} \propto \int\!\!\!\int T\left( {x,y} \right){I_i}\left( {x,y} \right)
\end{equation}
where $T$($x$, $y$) is the transmission function of the object.

\section{Experiment and results}

\begin{figure}[!ht]
  \centering
  \includegraphics[width=1\columnwidth]{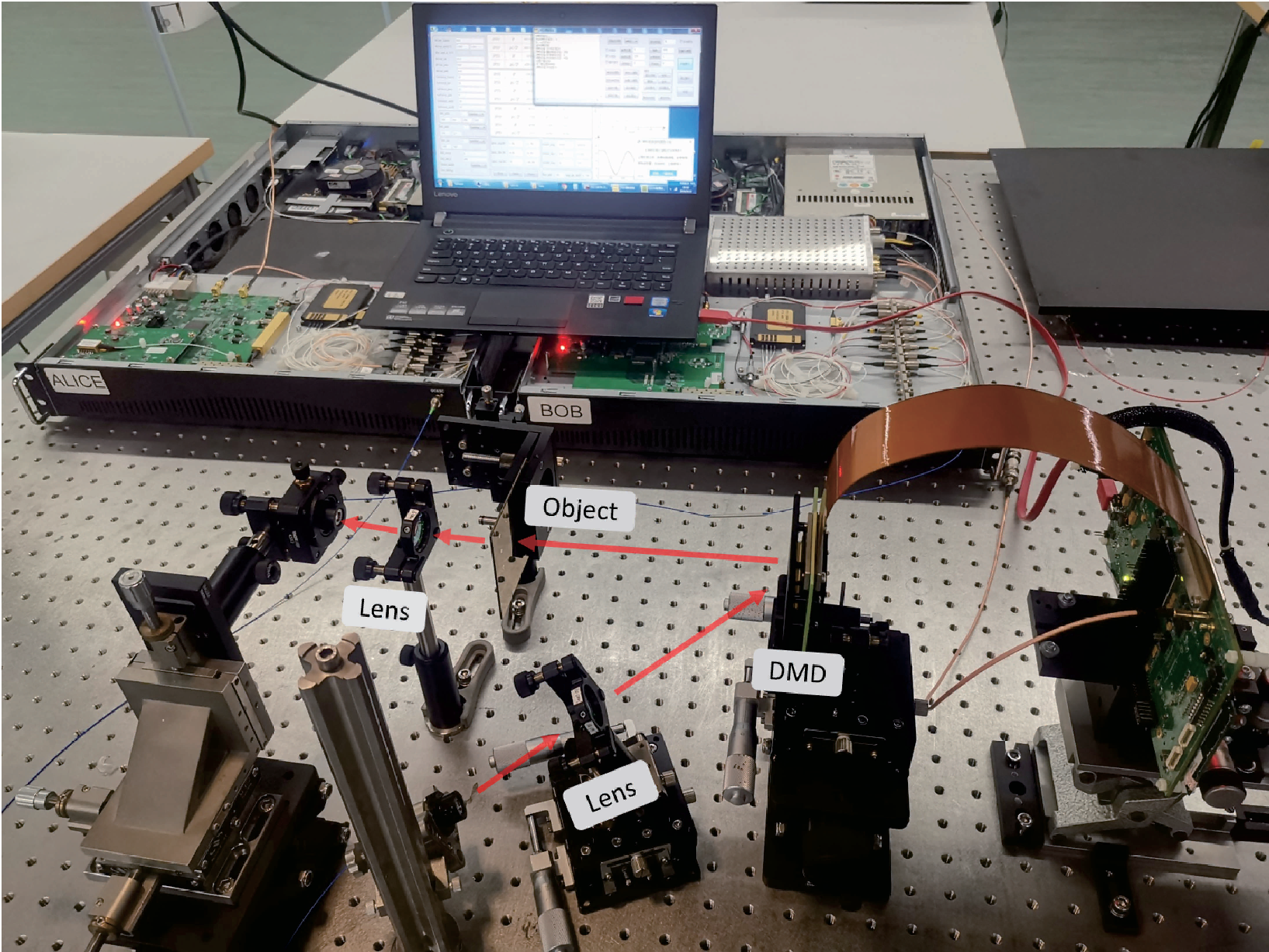}\\
  \caption{(color online) Schematic experimental setup of QSI system. The red arrows indicate the path of light roughly.}\label{exp}
\end{figure}

\begin{figure}[!ht]
  \centering
  \includegraphics[width=0.3\linewidth]{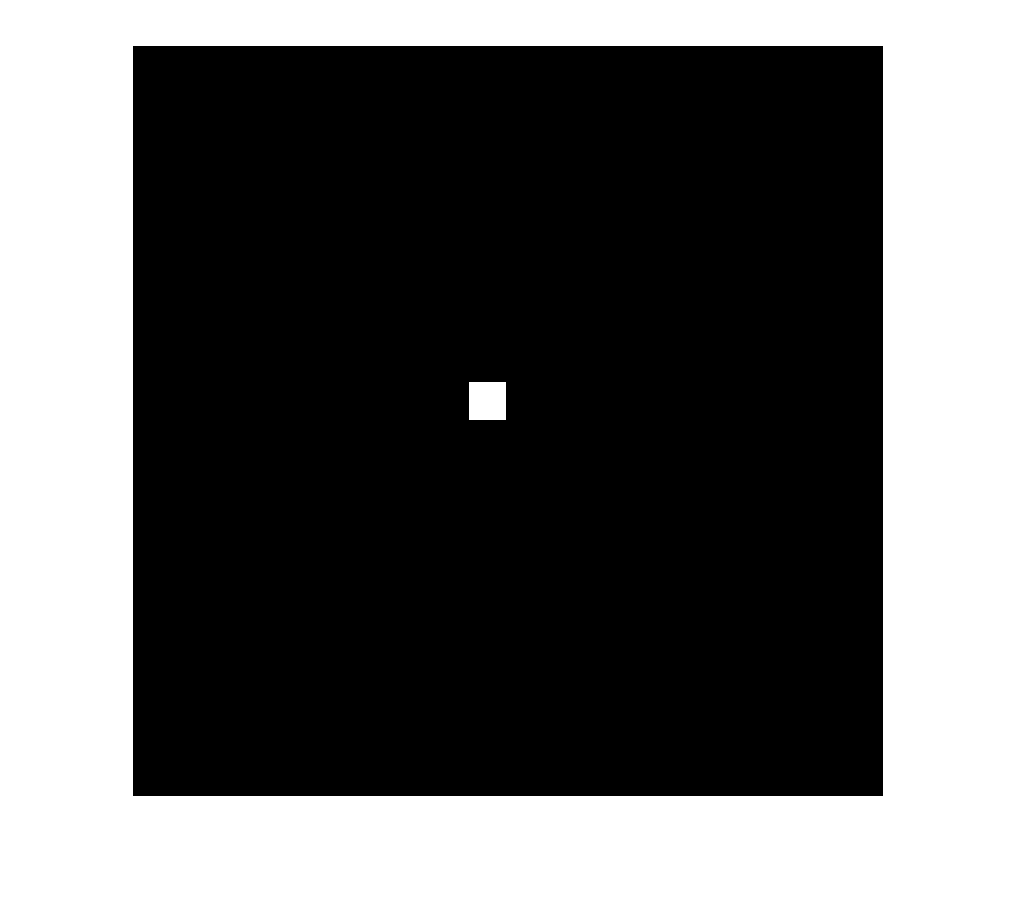}\\
  \caption{ The simplest binary pattern loaded into the DMD that only one block is in the “on” state and others are in the “off” state.}\label{binary_block}
\end{figure}

QSI system is set up as shown in the Fig. \ref{exp}. In the QKD system, the pulsed laser is working at a frequency of 40 MHz and a central wavelength of 1550.12 nm. The SPDs is working at Geiger mode and has a detection efficiency of 10$\%$ for photons. The intensities of signal, decoy and vacuum state are set to $\mu=$ 0.68, $\nu=$ 0.18 and 0 with the probabilities of 13$/$16, 2$/$16 and 1$/$16, respectively. In the quantum imaging part, the array of the DMD micro-mirrors (G4100 from TI, the control board is made by x-digit) is 1024 $\times$ 768, and the size of every micro-mirror (one pixel) is 13.68 $\mu$m $\times$ 13.68 $\mu$m. Since each micro-mirror is so small that the laser energy reflected from it is extremely low, a new array of 24 $\times$ 24 pixels is reselected and controlled with the same state in any case. The size of the new block is approximately 328.32 $\mu$m $\times$ 328.32 $\mu$m when ignore the gaps between the micro-mirrors. We choose only the square areas wherever covered by the light spot and the intensity is adequate enough. So, the DMD is divided into 20 $\times$ 20 $=$ 400 blocks,  In the experiment, Fig. \ref{binary_block} is the simplest binary pattern loaded into the DMD that only one block is in the “on” state and others are in the “off” state. In all the binary patterns, each block appears just only once, which is similar to spot scanning. In this way, only 400 binary patterns are needed, which can reduce the complexity of the imaging algorithm and greatly save the imaging time.

\begin{figure}[!ht]
	\centering
	\begin{tabular}{ccc}
		\includegraphics[width=0.3\linewidth]{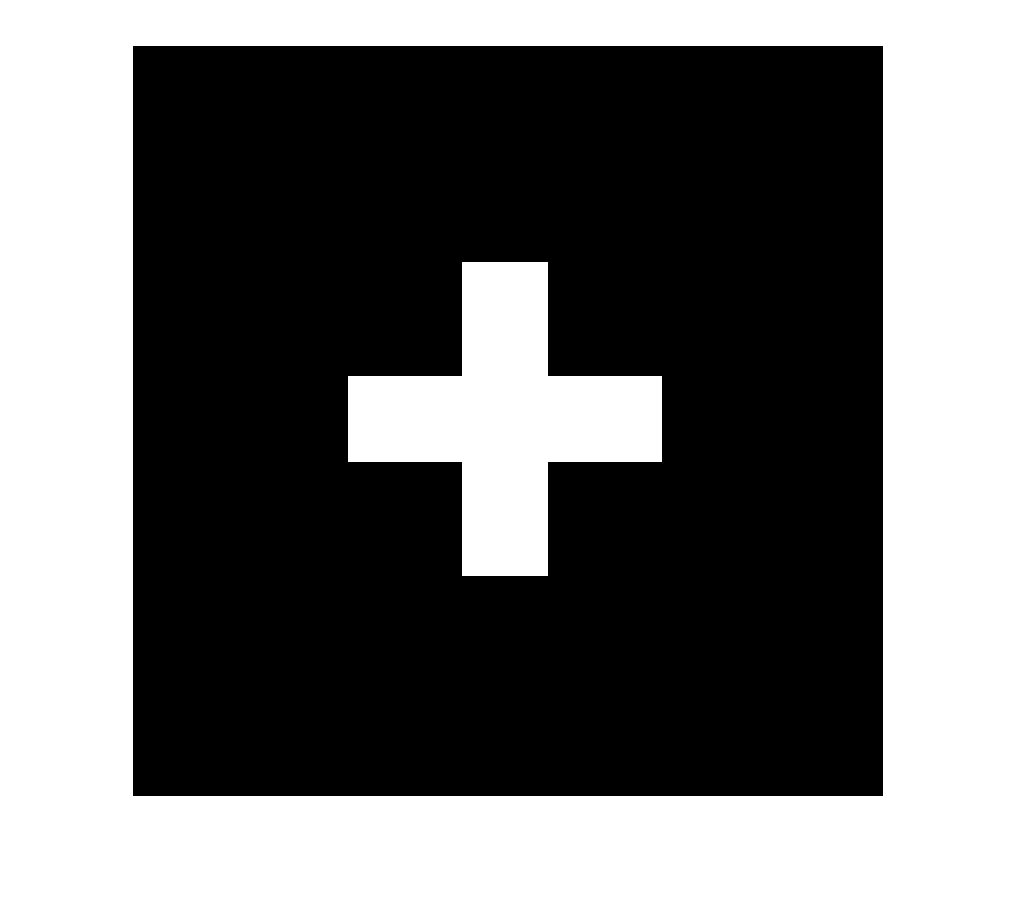}  &
		\includegraphics[width=0.3\linewidth]{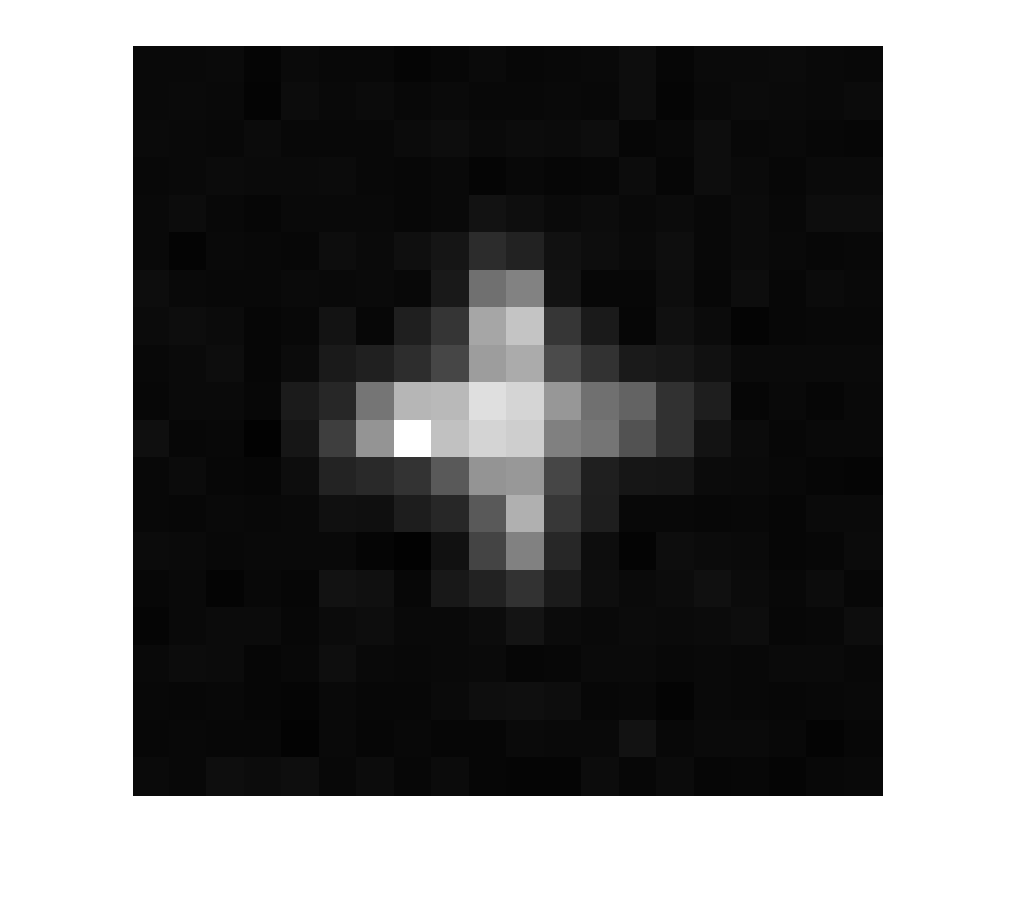}   &
		\includegraphics[width=0.3\linewidth]{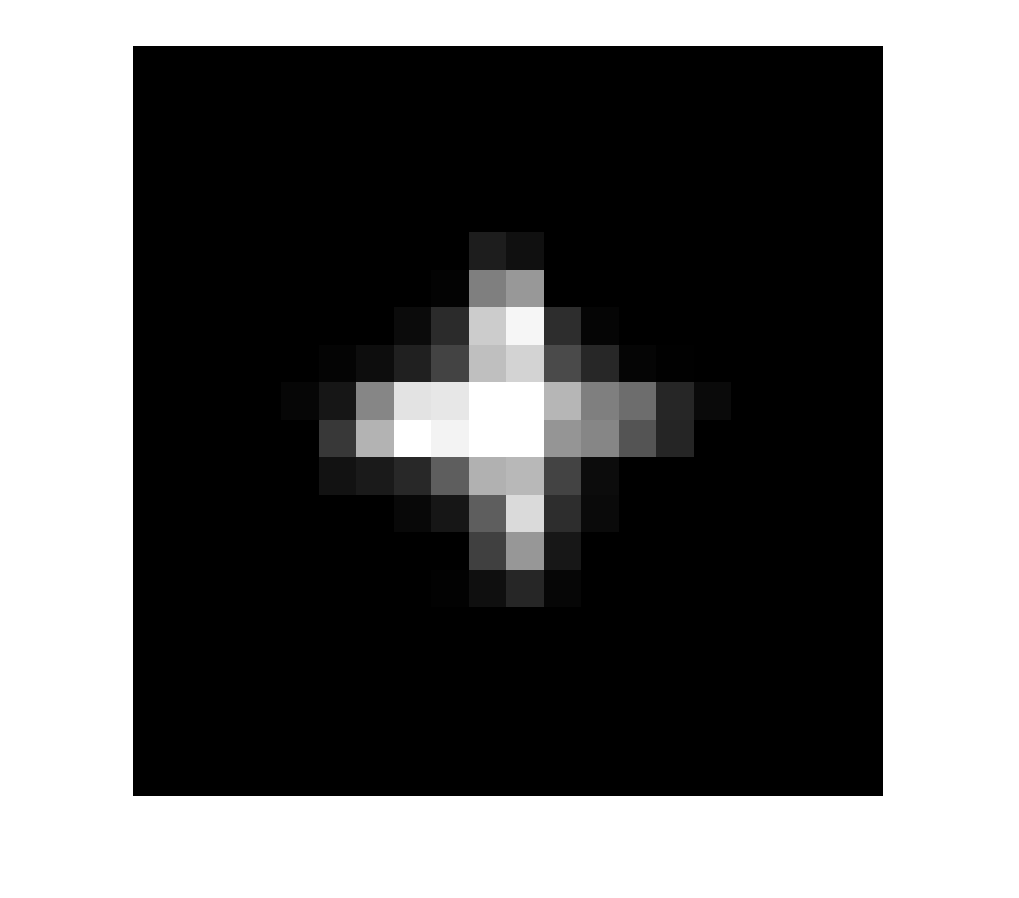} \\
		(a) & (b) & (c)\\
	\end{tabular}
	\caption{Results of numerical simulation and QSI experiment of a “+” type object. (a) is the result of numerical simulation of imaging of a “+” type object according to the Eq. (\ref{Oxy}). (b) is the result of using the measured signal intensity distribution for imaging of a “+” type object directly. (c) is the result of calculating the measured signal intensity and the presetting intensity distribution for imaging of a “+” type object according to the Eq. (\ref{Oxy}), which is better than (b).}
	\label{Fig:4}
	\vspace{-0.5em}
\label{image}
\end{figure}

After completing the connection of experimental device as shown in Fig. \ref{exp}, setting up the frames number $N =$ 400, frame frequency $f =$ 200 Hz. Running the DMD in the fully open mode, and making the QKD work normally through scanning and setting up every parameter within QKD before placing the object. Then, putting the “+” type object into this system, and loading 400 random binary patterns into the DMD. Running the whole system synchronously, and signal photon counts and QBER were measured during the display time of each pattern frame. The corresponding results of QKD procedure are shown in Table. \ref{table}, which are obtained by following the standard decoy-state BB84 QKD protocol. According to the Eq. (\ref{Oxy}), the reconstructed image of the object is shown in Fig. \ref{image}. The SNR is 23 dB, where SNR is defined as SNR $= {{{{\bar s}^2}} \mathord{\left/
 {\vphantom {{{{\bar s}^2}} {{\sigma _n}}}} \right.  \kern-\nulldelimiterspace} {{\sigma _n}}}$, $\bar s$ is average of the signal intensity, and $\sigma _n$ is the variance of the background intensity. The difference between the average intensity of the bright and dark regions of images is regarded as the signal, and the variation of dark background is considered as the noise.\upcite{q,cc} The lower bound of error rate is calculated to be 14.51$\%$ according to Eq.(\ref{etL}), and a secure key rate of 571.0 bps is achieved during the imaging period.

\begin{table}[hbp]
  \centering
  \caption{Experimental results of QKD procedure.}
\begin{tabular}{ccccc}
  \hline \hline
  $Q_{\mu}$ & $Q_{\nu}$ & $Y_0$ & $E_{\mu}$ & $E_{\nu}$ \\ \hline
  $2.69 \times 10^{-4}$ & $7.32 \times 10^{-5}$ & $3.0 \times 10^{-6}$ & $2.13 \%$ & $3.99\%$ \\
  \hline \hline
  \end{tabular}
 \label{table}
\end{table}

As can be seen from the Table. \ref{table} and Fig. \ref{image}, the image information of the object can be obtained by using the QSI scheme, and the average QBER is 3.99$\%$, which is below the lower bound and is proved that the imaging process is safe and real. Once the jammer of object tries to carry out the intercept-resend strategy, it is bound to lead to the over threshold increase of QBER under the premise of presence of photon counts. In general, we think that the jammer of object has carried on the intercept-resend strategy and the imaging is no longer safe and real when the average of QBER is more than 14.51$\%$. Therefore, our QSI system can be used as quantum secure radar against jammers.

But at the same time, we can see that the image resolution is not high. The main reason is the serious diffraction effect of the DMD. As we know, the DMD is composed of many micro mirrors, which is similar to a two-dimensional diffraction grating. It has a strong diffraction effect on monochromatic light, and becomes stronger with the increase of wavelength,\upcite{z} where can hardly even be used for imaging in the mid-infrared band. Therefore, even if all the micro mirrors of the DMD keep in the “off” state, the detector will still measure much energy due to the diffraction effect, which will greatly reduce the SNR and the imaging quality. As a result, neither the new block of the binary patterns can be too small, otherwise the diffraction noise will drown out the signal energy that make the SNR relatively low. Nor the new block can be too large, otherwise the imaging resolution will be dissatisfactory. Therefore, it is necessary to find a balanced block in the experiment that make the imaging quality best. In our experiment, the block with size of 328.32 $\mu$m $\times$ 328.32 $\mu$m is relatively appropriate. In addition, we can minimize the diffraction effect by the following three methods. Firstly, we can change the distribution of the diffraction spots and energy by adjusting the incident angle of light on the DMD to minimize the diffraction effect. Secondly, we can increase the distance between the DMD and the detector to reduce the diffraction effect, because the diffraction spots are distributed at some certain angles. The farther away the distance, the larger deviation of the diffraction spots, and the less stray light coupled into the detector. finally, we can design the wavelength of the light to match the DMD to reduce the diffraction effect, but all the devices need to be matched to this wavelength, such as the DMD, the laser and the detector of QKD, which is a troublesome and complex thing. In the future works, we will focus on the research how to reduce the diffraction effect of DMD to make the imaging quality optimal.

\section{Conclusion}
In conclusion, we proposed a QSI scheme based on the phase encoding and weak$+$vacuum decoy-state BB84 protocol QKD that can resist the intercept-resend strategy of the jammers. The authenticity and safety of the imaging process are ensured by the theoretical unconditional security of QKD, and the QBER and the secure key rate analytical functions of QKD are used to estimate whether the jammer of object carries out the intercept-resend attacks. In our experiment, we obtain a secure key rate of 571.0 bps and a secure average QBER of 3.99$\%$, which is below the lower bound of 14.51$\%$. In the imaging system, we synthetically use the CGI scheme, SPDs and DMD both with faster response speed that greatly improve the imaging efficiency. This scheme is suitable for the objects both with transmission or reflection. Furthermore, the system uses a WCS with invisible wavelength of 1550 nm, that can realize weak-light imaging and be immune to the stray light or air turbulence. Thus, it will become a better choice for quantum secure radar against jammers. But at the same time, it is undeniable that the more obvious diffraction effect of the DMD on the longer wavelength light will reduce the imaging SNR and quality. We believe that the diffraction effect can be minimized and the imaging quality can be improved by continuously adjusting the incident angle of light on the DMD, increasing the distance between the DMD and the detector or designing the wavelength of the light to match the DMD.

\section{Acknowledgment}
We acknowledge Xiaoming Lu, Di Jiang and Bin Chen for fruitful discussions and software support.

\end{document}